\documentstyle[aps]{revtex}
\input epsf

\newcommand{\degree}{\mbox{$^{\circ}$}}

\begin{document}

\title{Damping rates of the atomic velocity in Sisyphus cooling}

\author{
Laurent Sanchez-Palencia, Michele Schiavoni, Fran\c cois-R\'egis Carminati,
Ferruccio Renzoni and Gilbert Grynberg
}

\address{Laboratoire Kastler Brossel, D\'epartement de Physique de l'Ecole
Normale Sup\'erieure, \\ 24, rue Lhomond, 75231, Paris Cedex 05, France}

\date{\today{}}

\maketitle

\begin{abstract}
We present a theoretical and experimental study of the damping process of 
the atomic velocity in Sisyphus cooling. The relaxation rates of the atomic 
kinetic temperature are determined for a 3D lin$\perp$lin optical lattice.  
We find that the damping rates of the atomic temperature depend linearly on 
the optical pumping rate, for a given depth of the potential wells. This is
at variance with the behavior of the friction coefficient as calculated
from the spatial diffusion coefficients within a model of Brownian motion.
The origin of this different behavior is identified by distinguishing the 
role of the trapped and traveling atoms.
\end{abstract}


\section{\uppercase{Introduction}}
The last decades have witnessed an impressive progress in laser cooling
techniques, and nowdays it is possible to prepare and manipulate very
cold and dense atomic samples \cite{nobel97,metcalf}. Sisyphus cooling
\cite{sisifo} represents a milestone in the history of laser cooling,
particularly because this new cooling process demonstrated that the Doppler
limit could be broken \cite{lett}. Furthermore, it is an important 
peculiarity of the
Sisyphus process that the cooling mechanism not only leads to very low
temperatures, but also to the strong localization of the atoms near the
minima of the optical potential associated to the dipole force. Using
appropriate laser configurations, it is then possible to design
multidimensional modulations of the optical potential such that the atoms
result spatially ordered in periodic arrays whose periodicity is
of the order of the laser wavelength, the so-called optical lattices
\cite{robi}. The possibility to prepare atoms in spatially ordered structures
turned out to be essential for several studies in nonlinear dynamics
\cite{raizen,stoc,stoc2}.

Since the early works \cite{sisifo}, Sisyphus cooling has been
extensively studied, both experimentally and theoretically, in monodimensional
as well as multidimensional optical lattices. The dependence of the
steady-state kinetic temperature on the lattice parameters has been
carefully examined \cite{gatzke,jersblad,ellmann}. It is by now a well
established result that the temperature scales as the depth of the optical
potential \cite{gatzke}. More recently, it has been shown that in a 3D 
lin$\perp$lin
configuration (see below) the temperature is on one hand independent of the
lattice constant \cite{epj1,lsp} and on the other hand anisotropic
\cite{lsp,jersblad2}.  Much less attention was payed to the study
of the time scale of the damping of the atomic velocity in an optical lattice.
So far, only {\it indirect} measurements of the cooling rate of atoms have been
presented \cite{raithel}. In that work, the cooling rates were inferred 
from the atomic localization rates measured by using Bragg scattering as 
a sensitive probe.

In this paper we report a theoretical and experimental study of the damping
process of the atomic velocity in Sisyphus cooling. More precisely, we
measure, both in the experiments and in the numerical simulations, the width
of the atomic velocity distribution (which is proportional to the square
root of the kinetic energy of the atomic sample) for different lattice 
durations within the thermalization phase. It has thus been possible to 
monitor the damping of the kinetic energy in a direct way. We show that the 
kinetic energy decreases exponentially during this phase and we determine the 
corresponding damping rate $\Gamma_T$ as a function of the lattice 
parameters (detuning from atomic resonance and intensity of the laser beams). 
We find that $\Gamma_T$ is proportional to the optical pumping rate and is 
strongly anisotropic.
We also compare the cooling rate $\Gamma_T$ to the friction coefficient 
$\alpha$ as deduced from the spatial diffusion coefficients by using the 
Einstein relation for Brownian motion. We find that these two quantities 
show completely different dependencies on the lattice parameters.
The origin of this different behavior is identified by distinguishing the
role of the trapped and traveling atoms.

The present investigation is motivated by a parallel study of
probe-transmission spectroscopy of atoms cooled in a dissipative optical 
lattice \cite{yanko,ray}. The properties of the optical lattice are 
studied by monitoring the transmission of a weak probe field through
the atomic sample as a function of the relative detuning between the probe and
the lattice fields. The knowledge of the relaxation rates 
of the different atomic observables (atomic density, atomic velocity, 
magnetization, etc) is necessary to understand the physical origin of the 
Rayleigh resonance that appears at the center of the probe transmission 
spectra. It is well established that the Rayleigh resonance originates from
the scattering of photons from the lattice beams on the grating of
an atomic observable induced by the pattern of interference between the
probe and the lattice beams, and that the width of the Rayleigh line
is given by the relaxation rate of this observable \cite{courtois96} .
However which atomic observable determines the Rayleigh line in an 
optical lattice is still an open question \cite{ray}. Hence, the knowledge
of the relaxation rates of the different atomic observables should allow,
through comparison with the width of the Rayleigh line, the
identification of the observable leading to the Rayleigh resonance.

\section{\uppercase{Experimental set-up}}

In our experiment $^{85}$Rb atoms are cooled and trapped in a 3D
dissipative optical lattice red detuned with respect to the
$F_g=3\to F_e=4$ D$_2$ line transition. The procedure to load the 
atoms in the optical lattice is the standard one used in previous 
experiments \cite{epj1}: the rubidium atoms are first cooled and 
trapped in a magneto-optical trap (MOT) for about $1$ s; the temperature 
of the atoms in the MOT is of the order of $200~\mathrm{\mu K}$. 
Then the MOT magnetic field and laser beams are turned off and the 
lattice beams are turned on by using acousto-optical modulators (AOM) 
with a fall-rise time shorter than 1 ${\mu}$s.

\begin{figure}[ht]
\begin{center}
\mbox{\epsfxsize 2.5in \epsfbox{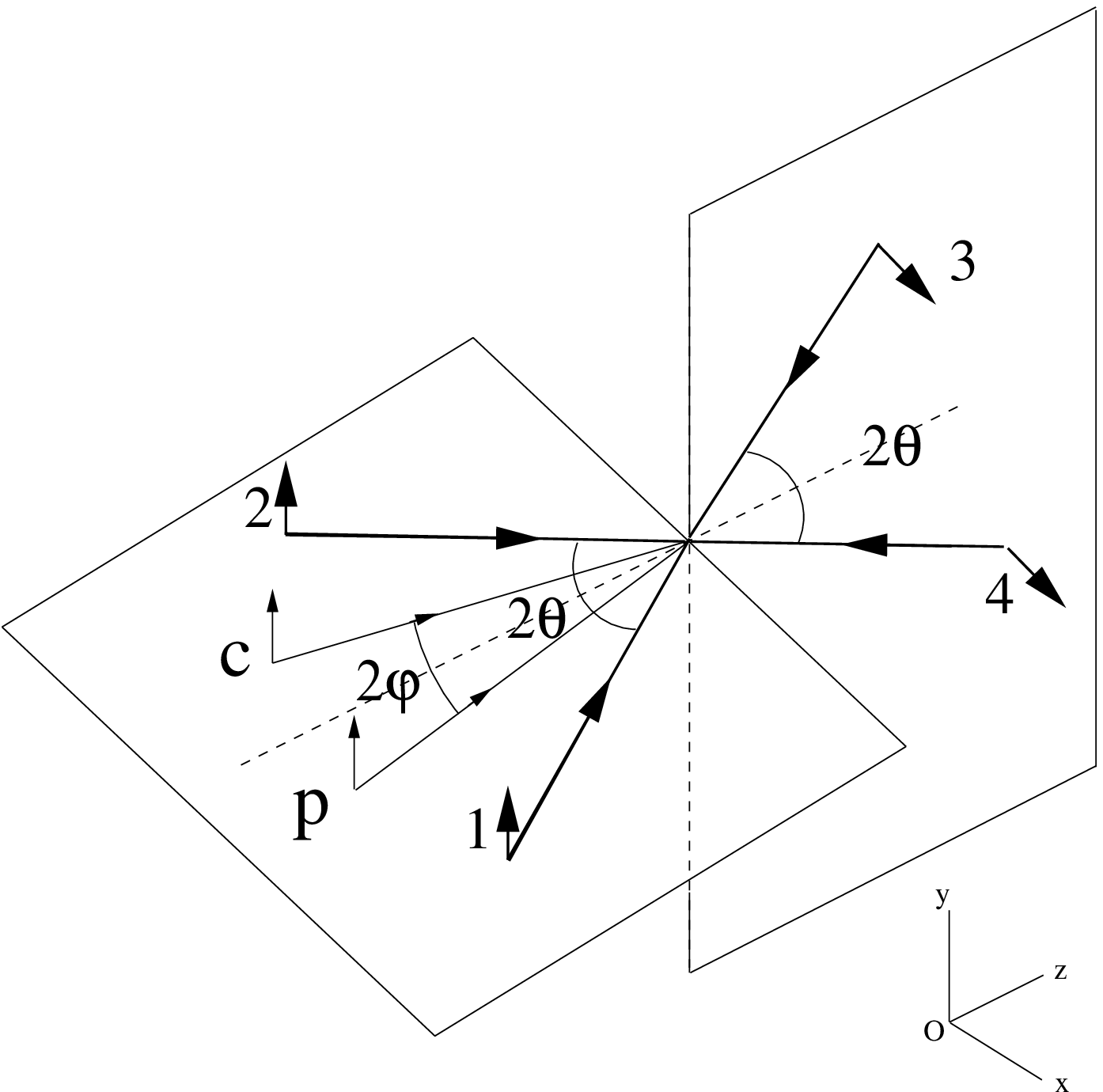}}
\end{center}
\caption{Laser fields configuration for the 3D-lin$\perp$lin optical lattice.
The beams $1-4$ generate the 3D periodic potential. For the measurement
presented in this work, the angles between the lattice beams are kept fixed
at $\theta=30^o$. Two additional laser beams (c and p), are introduced for
the temperature measurement. They are both linearly polarized along the
$y$-axis. The angle between beams c and p is $2\varphi = 25\degree$.}
\label{fig1}
\end{figure}

The three-dimensional  periodic optical lattice is generated by the
interference of four linearly polarized laser beams, arranged in the
3D lin$\perp$lin configuration \cite{robi} (Fig.~\ref{fig1}): two
$y$-polarized beams propagate in the $(xOz)$ plane with a relative angle
$2\theta$, while two $x$-polarized beams propagate in the $(yOz)$ plane
and form also an angle $2\theta$. This arrangement results in a periodic
modulation of the light polarization and light intensity, which produces
a periodic modulation of the light shifts of the different atomic
ground-state sublevels (optical potentials). The optical potential has
minima located on a orthorombic lattice and associated with pure circular
(alternatively $\sigma^{+}$ and $\sigma^{-}$) polarization. The lattice
constants, i.e. the distance (along a major axis) between two sites of
equal circular polarization are $\lambda_{x,y}=\lambda/\sin\theta$ and
$\lambda_z =\lambda/(2\cos\theta)$, where $\lambda$ is the laser field 
wavelength.

This configuration is well known to be suited for Sisyphus cooling \cite{robi}
for which the steady-state kinetic energy of the atomic sample results from 
the competition
of cooling and heating processes. The optical pumping between the different
sublevels of the atomic ground state combined with the spatial modulation of
the light shifts leads to the cooling of atoms, while the heating is essentialy
due to the statistical fluctuations of the dipole force \cite{sisifo}. For deep enough potential
modulations, Sisyphus cooling leads to atomic kinetic energies lower than the
potential depth so that the atoms get trapped at the minima of the optical
potential wells, thus producing a periodic array of atoms.

The temperature of the atoms is measured with the method of recoil-induced
resonances \cite{courtois94,meacher,chi,mileti}. The atoms are released
from the lattice by turning off the lattice beams and two new laser beams
(the pump (c) and the probe (p) beams, see Fig.~\ref{fig1}) are 
introduced. Beams c and p are detuned by $49$ MHz from atomic resonance 
and their intensities are of the order of 100 $\mu$W/cm$^2$.
They cross the atomic sample in the $(xOz)$ plane, and they are symmetrically
displaced with respect to the $z-$ axis forming an angle of 
$2\varphi = 25\degree$. The diamater of the pump beam is about 1 cm, 
larger than the atomic cloud, while the probe beam is very thin (less than
500 ${\mu}$m in diameter). The frequency of the probe beam is scanned 
around the pump frequency in 1 ms. We notice that for these parameters of
the pump and probe beams, the heating due to their absorption is negligible.
The probe transmission is monitored as a function of the detuning $\delta$ 
between the pump
and probe fields. For equal pump and probe polarizations, the ground states
$|\alpha, p\rangle$ and $|\alpha, p'\rangle$ with the same internal
quantum number $\alpha$ but different atomic momenta are coupled by
Raman transitions. The difference in population of these states determines
the absorption or gain of the probe, and the probe transmission spectrum
is thus proportional to the derivative of the atomic velocity distribution 
\cite{meacher}. These distributions were found to fit well to Gaussian functions
so  that it is straightforward to calculate the mean kinetic energy $E_K$ in 
the $j$- direction and to derive an atomic kinetic temperature
\begin{equation}
E_{K_j} = \frac{k_B T_j}{2}
\label{temperature}
\end{equation}
from the width  of the resonance in the probe transmission spectrum.
The geometry of our pump and probe fields corresponds to
measurement of the temperature in the $x-$ direction.

\section{\uppercase{Experimental results}}

\begin{figure}[ht]
\begin{center}
\mbox{\epsfxsize 3.in \epsfbox{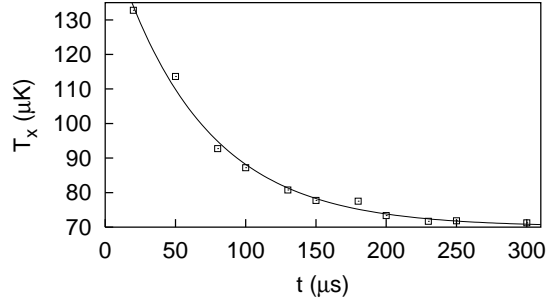}}
\end{center}
\caption{Experimental results for the temperature $T_x$ in the $x$-direction
of the atoms cooled
in the optical lattice as a function of the duration $t$ of the lattice.
The lattice detuning is $\Delta = 48$ MHz, the intensity per lattice beam
$I_L=4.5$ mW/cm$^2$. The solid line is the fit to the exponentially
decreasing function Eq.~(\ref{relaxation}).}
\label{fig2}
\end{figure}

In the experiment, the duration of the optical lattice is carefully
controlled and can be varied in steps of $10 \mu$s. We 
measure the kinetic temperature of the atomic cloud as a function of
the duration $t$ of the optical lattice for different lattice parameters
(intensity per lattice beam $I_L$ and lattice detuning $\Delta$ from
atomic resonance).  We plot in Fig.~\ref{fig2} a typical variation of
the measured kinetic temperature $T_x$ as a function of the time $t$ together
with a fit of the data to a function of the form 
\begin{equation}
T_x(t)=T_{f_x}+ (T_i-T_{f_x}) \exp(-\Gamma_{T_x}\cdot t).
\label{relaxation}
\end{equation}

The kinetic temperature decreases exponentially with the lattice duration 
$t$ towards a steady-state temperature, and the fit to Eq.~(\ref{relaxation})
provides the steady-state temperature $T_{f_x}$ as well as the damping rate
$\Gamma_{T_x}$ of the atomic temperature. From experimental data as those
of Fig.~\ref{fig2}, we determined $\Gamma_{T_x}$ and $T_{f_x}$ in a broad
range of lattice parameters (intensity per lattice beam $I_L$ and
lattice detuning $\Delta$). We found that the temperature shows a linear
increase with $I_L$ and decreases against $\Delta$, in agreement with earlier
studies. A more complete account of the dependence of the steady-state
temperature on the lattice parameters has already been reported in
Ref.~\cite{epj1} and will not be repeated here. The novel result of the 
present work consists instead in the determination of the damping rate of 
the temperature.

\begin{figure}[ht]
\begin{center}
\mbox{\epsfxsize 3.in \epsfbox{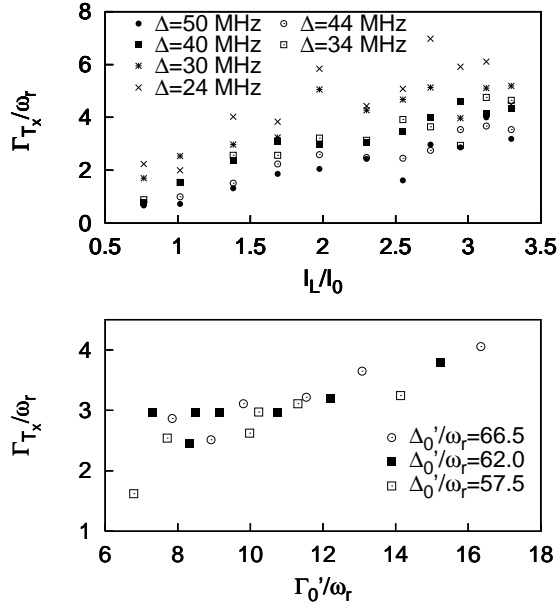}}
\end{center}
\caption{Experimental results for the damping rate $\Gamma_{T_x}$ of the
atomic temperature in the $x-$ direction. In the upper graph $\Gamma_{T_x}$
is plotted as a function of the intensity per lattice beam $I_L$, for
different values of the lattice detuning $\Delta$. In the lower graph
$\Gamma_{T_x}$ is plotted as a function of the optical pumping rate
$\Gamma_0'$, for different values of the light shift per beam $\Delta_0'$.
Here $\omega_r$ is the atomic recoil frequency, and $I_0$ the saturation
intensity.}
\label{fig3}
\end{figure}

Experimental results for $\Gamma_{T_x}$ are reported in Fig.~\ref{fig3} as
a function of the lattice beam intensity at fixed lattice detuning (top)
and as a function of the optical pumping rate $\Gamma_0'\propto I_L/\Delta^2$
at fixed light shift per beam $\Delta_0'\propto I_L/\Delta$ (bottom).
We find that $\Gamma_{T_x}$ ranges from a few kHz to a few tens of kHz for
typical intensity and detuning of the lattice beams. More precisely, from
the experimental data, we identify the following general behavior of the
damping rate as a function of the lattice parameters: $\Gamma_{T_x}$ is an
increasing function of the lattice beams intensity $I_L$, the dependence
being linear within the experimental error, and a decreasing function of
the lattice detuning $|\Delta|$. When plotted against the optical pumping rate,
$\Gamma_{T_x}$ is found to be proportional to $\Gamma_0'$, and almost
independent of the light shift per beam. The observed behavior is in 
good agreement with the results of Raithel {\it et al} \cite{raithel},
and will be interpreted with the help of Monte-Carlo simulations.

\section{\uppercase{Theoretical analysis}}

We analyze theoretically the damping process of the atomic velocity with
the help of semi-classical Monte-Carlo simulations. Taking advantage of the
symmetry between the $x-$ and $y-$ directions (see Fig.~\ref{fig1}) we
restricted the atomic dynamics in the $(xOz)$ plane. The calculations are made
for a $J_g=1/2\to J_e=3/2$ atomic transition, as customary done in numerical
analysis of Sisyphus cooling. The numerical calculations in this simplified
configuration are expected to give the same dependencies of the dynamical 
quantities on the lattice parameters as full 3D calculations, so they are 
useful to study the main feautures of the cooling dynamics. However, a
direct quantitative comparison to the experimental results is not necessarily
meaningful because the restriction of the dynamics to two dimensions may 
introduce scaling factors \cite{lsp}.

\subsection{Cooling rates}

\begin{figure*}[ht]
\begin{center}
\mbox{\epsfxsize 3.25in \epsfbox{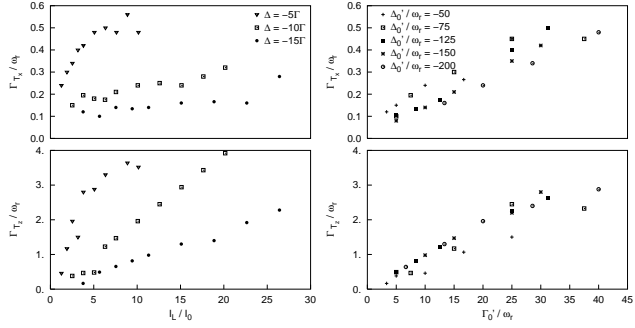}}
\end{center}
\caption{Numerical results for the damping rates of the atomic temperature
in the $x-$ and $z-$ directions. In the left hand column the damping rates 
are plotted as functions of the intensity per lattice beam $I_L$, for 
different values of the lattice detuning $\Delta$. In the right hand column, 
the same data are reported as functions of the optical pumping rate 
$\Gamma_0'$, for different values of the light shift per beam $\Delta_0'$. 
Here $\Gamma$ is the width of the excited state.}
\label{fig4}
\end{figure*}

We first determine the mean kinetic energies along the $x-$ and $z-$ 
directions as functions of time in the thermalization phase. The mean
kinetic energy is obtained by averaging over about 5000 trajectories.
For both directions, the temperature shows an exponential decrease with
the lattice duration. By fitting the temperature $T_x$ and $T_z$ with 
a function of the form (\ref{relaxation}), we determine the 
corresponding relaxation rates $\Gamma_{T_j}$ in the $x-$ and $z-$ 
directions, with results as in  Fig.~\ref{fig4}.

We notice that the damping rates for the two orthogonal directions are 
very different, with $\Gamma_{T_x}$ typically about one order of magnitude 
smaller than $\Gamma_{T_z}$. This property certainly relates to the strong 
anisotropies of the optical potential and of the optical pumping rates in the
3D lin$\perp$lin lattice. 
%
% We verified that this anysotropy is not purely due to the different lattice 
% periods in the $x-$ and $z-$ directions, i.e. the anysotropy cannot be 
% reabsorbed by a rescaling of the $x$ and $z$ coordinates by the 
% corresponding lattice periods.  
%
Previous work already
pointed out anisotropic temperatures \cite{lsp,jersblad2} and anisotropic 
spatial diffusion coefficients \cite{epj1,lsp} for a 3D-lin$\perp$lin lattice.

The graphs of the left-hand column of Fig.~\ref{fig4} report the results for 
the damping rates $\Gamma_{T_j}$ as functions of the intensity per lattice 
beam $I_L$ for different values of the lattice detuning $\Delta$ and confirm 
the general behavior observed in the experiment: $\Gamma_{T_j}$ is an
increasing function of the lattice intensity, and a decreasing function of
the lattice detuning.

On the right-hand column of Fig.~\ref{fig4} the numerical results for
the damping rates are reported as functions of the optical pumping rate
per lattice beam $\Gamma_0'$ for fixed values of the light shift per beam
$\Delta_0'$, i.e. for fixed values of the depth of the potential wells. 
This allows a more transparent interpretation of our results. Indeed, the
damping rates are found to be proportional to the optical pumping rate in 
a large range of light shifts per lattice beam within a numerical error of 
about $10~\%$.  This is in qualitative agreement with our experimental 
observations, and also with the indirect measurements of the cooling rates
of Raithel {\it et al} \cite{raithel}.

\subsection{Friction coefficients}

\begin{figure}[ht]
\begin{center}
\mbox{\epsfxsize 2.5in \epsfbox{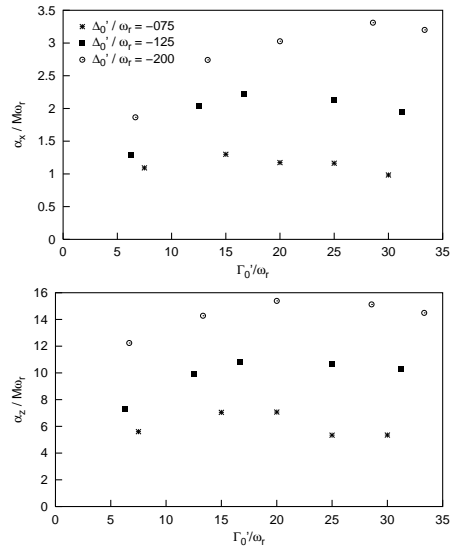}}
\end{center}
\caption{Numerical results for the friction coefficients in the $x-$ and 
$z-$ directions as functions of the optical pumping rate $\Gamma_0'$ for 
various values of the light shift per beam $\Delta_0'$.}
\label{fig5}
\end{figure}

It is interesting to compare the results obtained for the damping rate of 
the kinetic energy with the well known friction coefficients for Sisyphus
cooling \cite{sisifo}. Such a coefficient $\alpha$ is introduced in the 
low-velocity limit where the mean cooling force has the form $F=-\alpha v$.
We determined the friction coefficients $\alpha_x$ and $\alpha_z$ for
the two orthogonal directions. A possible way to calculate the friction
coefficients is to introduce an arbitrary constant force $F_j$ along
the direction $j$. The mean friction force compensates this force, with 
the atomic cloud set into a motion along $j$ at a constant mean velocity
$\left\langle v_j \right\rangle = F_j / \alpha_j$. The friction 
coefficient $\alpha_j$ is then easily derived by calculating the velocity
$\left\langle v_j \right\rangle$ for various $F_j$.
A different approach to the calculation of the friction coefficients
is based on the assumption that the atomic dynamics can be described 
as a Brownian motion. Under this assumption, the friction coefficients
can be derived from the steady-state kinetic temperature $T_j$ and the 
spatial diffusion coefficient $D_{s_j}$ in the $j$-direction by using the 
Einstein relation for Brownian motion $\alpha_j = k_B T_j / D_{s_j}$.
In previous work \cite{lsp}, these two approaches have been compared and 
shown to lead to the same values for the friction coefficients.  
Also for the present choice of interaction parameters, we found that 
the two methods lead to values of $\alpha_j$ in 
excellent agreement \cite{remark}. These values are shown in Fig.~\ref{fig5} 
as functions of $\Gamma_0'$ for different values of the light shift 
per beam $\Delta_0'$. We 
notice that also for the friction coefficients there is a strong 
anisotropy, with  $\alpha_x$  about $5$ times smaller than $\alpha_z$. 
The results for the friction coefficients have to be compared with the 
values of $\Gamma_{T_x}$ and $\Gamma_{T_z}$ shown in Fig.~\ref{fig4}.
The values $\alpha_j/M$ are about one order of magnitude larger
than those for $\Gamma_{T_j}$. Furthermore, the two quantities do not show
the same dependence on the lattice parameters. In fact,  $\alpha_j$ weakly 
depends on $\Gamma_0'$ but strongly depends on $\Delta_0'$ in the range of 
parameters considered here. This behaviour is completely different from 
the one for $\Gamma_{T_j}$ in the same range of lattice parameters. 

\subsection{Discussion}

The very different dependence on the lattice parameters shown by the 
damping rates of the kinetic energy and the friction coefficients 
is rather surprising. In fact, the description of the atomic dynamics 
as a Brownian motion suggests that the friction coefficient in a given
direction, rescaled by the atomic mass $M$, is simply the half of the 
damping rate of the kinetic energy 
in the same direction. The different behavior shown by these two 
quantities can be explained by distinguishing the role of the trapped 
and traveling atoms. In fact both in the thermalization phase and at
thermal equilibrium there are high energy atoms (h-e) that travel 
over the lattice sites and low energy atoms (l-e) that are trapped 
in the potential wells. As discussed in detail in Ref.~\cite{lsp}, 
the high energy atoms experience Sisyphus cooling, undergo a Brownian 
motion, and their dynamics can be characterized by a friction coefficient
$\alpha_j$. Since these atoms undergo a Brownian motion one may expect
\begin{equation}
\left(\frac{\alpha_j}{M}\right)_{h-e}=\left(\frac{\Gamma_{T_j}}{2}\right)_{h-e}
\end{equation}

On the other hand, low-energy atoms are trapped in the potential wells
and do not undergo a Brownian motion. Therefore they show a very large 
(ideally divergent) friction coefficient. In fact, the application of a 
weak force $F_j$ does not set them into a uniform motion, but only changes 
their center of oscillation in the wells. Therefore for these atoms the 
friction coefficient $\alpha_j=F_j/\langle v_j\rangle$ is divergent. 
Furthermore, trapped atoms do not loose kinetic energy by Sisyphus cooling
and therefore they are characterized by a cooling rate 
$\Gamma_{T}$ ideally equal to zero. It follows that for the low-energy 
trapped atoms
\begin{equation}
\left( \frac{\alpha_j}{M}\right)_{l-e} \gg
\left( \frac{\Gamma_{T_j}}{2}\right)_{l-e} ~.
\end{equation}
It is then clear that the presence of trapped atoms invalidates the simple
relation $\alpha_j/M=\Gamma_{T_j}/2$ and lead to a cooling rate $\Gamma_{T_j}$
smaller than $2\alpha_j/M$, as observed in our numerical simulations.

\section{\uppercase{Conclusions}}

In summary, we have determined experimentally the damping rates of the 
atomic kinetic energy (kinetic temperature) in Sisyphus cooling, and have 
examined their general dependencies on the lattice parameters.
We have also made
numerical semi-classical Monte-Carlo simulations. We have shown that the 
damping rates of the atomic kinetic temperature depend linearly on the 
optical pumping rate and are strongly anisotropic. We have also determined
the friction coefficients from the atomic spatial diffusion coefficinets 
and we have
shown that they weakly depend on the optical pumping rate but strongly 
depend on the potential depth. The very different behavior of the damping 
rate of the kinetic temperature $\Gamma_{T_j}$ and of the friction
coefficient $\alpha_j$ has been explained by distinguishing the role of 
the trapped and traveling atoms.

The presented results for the damping rates of the atomic kinetic temperature 
in Sisyphus cooling are useful for the analysis of the Rayleigh resonance 
detected in probe-transmission spectroscopy of atoms cooled in dissipative 
optical
lattices \cite{yanko,ray}. Indeed, the knowledge of the damping rates of the
different atomic observables excited in the optical process is necessary to
determine the physical origin of the resonance. Preliminary measurements 
for the probe transmission show 
that the width of the Rayleigh resonance is of the same order of magnitude of
the value for $\Gamma_{T_j}$ presented in this work. This suggests that the
atomic velocity is one of the observables which, excited in the optical 
process determined by the probe and lattice fields, contribute to the Rayleigh
resonance. Further work is however necessary for a definitive conclusion.

\acknowledgments

Laboratoire Kastler Brossel is an "unit\'e mixte de recherche de l'Ecole
Normale Sup\'erieure et de l'Universit\'e Pierre et Marie Curie associ\'ee
au Centre National de la Recherche Scientifique (CNRS)".

\newpage

\end{document}